\documentclass[pra,aps, showpacs, twocolumn, a4paper,tightenlines,balancelastpage]{revtex4}
\usepackage{amsmath,amsfonts,amssymb,mathrsfs,bm}
\usepackage{verbatim,psfrag,times,CJK}
\usepackage{graphicx,graphics,color,epsfig}
\usepackage{flafter}

\usepackage{indentfirst}
\begin{document}

\newcommand{\ket}[1]{| #1 \rangle}
\newcommand{\bra}[1]{\langle #1 |}

\title{Entanglement created by spontaneously generated coherence}

\author{Zhao-hong \surname{Tang}$^{a,b}$}
\author{Gao-xiang \surname{Li}$^{a}$}
\email{gaox@phy.ccnu.edu.cn}
\author{Zbigniew \surname{Ficek}$^{c}$}

\affiliation{$^{a}$Department of Physics, Huazhong Normal University, Wuhan 430079,
PR China\\
$^{b}$School of Science, Wuhan Institute of Technology, Wuhan 430073, PR China\\
$^{c}$The National Centre for Mathematics and Physics, KACST, P.O. Box 6086, Riyadh 11442, Saudi Arabia}

\begin{abstract}
We propose a scheme able to generate on demand a steady-state entanglement between two non-degenerate cavity modes. The scheme relies on the interaction of the cavity modes with driven two or three-level atoms which act as a coupler to build entanglement between the modes.  We show that in the limit of a strong driving, crucial for the generation of entanglement between the modes is to imbalance populations of the dressed states of the driven atomic transition. In the case of a three-level V-type atom, we find that a stationary entanglement can be created on demand by tuning the Rabi frequency of the driving field to the difference between the atomic transition frequencies. The resulting degeneracy of the energy levels together with the spontaneously generated coherence generates a steady-state entanglement between the cavity modes. It is shown that the condition for the maximal entanglement coincides with the collapse of the atomic system into a pure trapping state. We also show that the creation of entanglement depends strongly on the mutual polarization of the transition atomic dipole moments.
\end{abstract}

\pacs{42.50.Dv, 42.50.Gy}

\maketitle

\section{Introduction}

The generation of continuous variable (CV) entangled light has attracted a significant interest due to a potential application in quantum information science, specifically in quantum teleportation~\cite{VL00}, quantum telecloning~\cite{VL01}, and quantum dense coding~\cite{JJ03}. Continuous variables offer the possibility to create entanglement deterministically and different nonlinear processes have been proposed to generate CV two-mode entangled beams~\cite{CH01,JV03,XH05,HT05,LFL10} including nondegenerate parametric down-conversion~\cite{ZY00,ZY92} and nondegenerate four-wave mixing processes~\cite{RE85,MM07,GS86,GX07,DL01}. Recently, the four-wave mixing process has been proposed as a potential source of narrow-band entangled beams, an important resource for quantum memory storage~\cite{BJ04} and long-distance communications~\cite{DLM00}.

Of particular interest for CV entanglement are cavity QED systems where entanglement between cavity modes can be created by coupling the modes to an atomic system or nonlinear crystal located inside the cavity~\cite{RM05,VB05,ZL06}. It was shown that for the generation of entanglement between cavity modes, it is essential to create a coherence in the coupling (or entangling) system. Typical systems for entangling the modes are multi-level atoms or nonlinear crystals where the coherence can be established initially by a preparation of the atoms in a linear superposition of their energy states or can be created dynamically by a suitable driving of the atoms through four-wave mixing~\cite{RE85,MM07,GS86,GX07,DL01} or Raman-type processes~\cite{xs05,kz07,qa09,lh09}.

The coherence is subjected to dissipation due to the decoherence process and over a long time it might be difficult to maintain the coherence large enough for entangling the cavity modes. The main source of decoherence is spontaneous emission resulting from the interaction of the atoms with the environment. On a microscopic scale, the spontaneous emission can be reduced or even completely eliminated, but it could be difficult to eliminate on a macroscopic scale where one would like to create entanglement using macroscopic atomic ensembles. This raises an important question of how to eliminate the decoherence or how to maintain a large coherence in the presence of the decoherence.

In this paper, we propose a system formed by a three-level atom located inside a two mode cavity that can generate the maximal stationary entanglement between the cavity modes in the presence of decoherence. The atom is modelled as a V-type system where the dipole allowed transitions can be independent of each other or can be correlated through the spontaneously generated coherence (SGC)~\cite{fs04}. The atom is driven by an external laser field coupled exclusively to only one of the atomic transitions. We use the dressed-atom approach and show that the effective three-level system of dressed states comprises a suitable medium for a non-linear coupling between the cavity modes. We work in the strong driving limit which assumes that the Rabi frequency of the laser field is much larger than the transition damping rates and the coupling strengths of the cavity modes to the atomic transitions. This prompts us to apply the secular approximation which ignores the coupling of the populations of the dressed states to the coherences. It is known that non-secular terms, although small can have a destructive effect on coherence effects~\cite{xs05,kz07} or may even have constructive effects and lead to interesting novel features~\cite{lb91,ss93,tl09}. However, we are interested in features created by the SGC rather than features created by the coherence induced by the driving field and therefore neglect the non-secular terms.

We consider four scenarios, where the cavity modes couple to the same or different atomic transitions that could be correlated or independent of each other. The first scenario represents a situation in which the atomic transitions are independent of each other and both cavity modes couple to the same atomic transition that, in addition, is driven by a strong and in general off-resonant laser field. Physically, this system behaves as a driven two-level system and the driving field occurs as a dressing field for the atoms. We demonstrate that the necessary and sufficient conditions for generation of the maximal entanglement between the modes is to create the complete population inversion between the dressed states of the coupling atomic system. A population difference between the dressed states occurs for an off-resonant driving field. Since for a strong driving field there is no coherence between the dressed states, one could conclude that the  entanglement occurs without coherence in this case. However, for a detuned driving field, a coherence actually occurs between the two bare states of the system. In other words, in the bare atom picture, the entanglement is created with coherence. We find that the maximal entanglement cannot be created in this scenario since it is not possible to create a large population difference between the dressed states and at the same time maintain a strong coupling between the cavity modes mediated by the atom.

In the second scenario, we include the coupling between the atomic transitions through the SGC, a close analog on the schemes of quantum-state engineering by dissipation~\cite{dm08,kb08,bz10,dt10,vw09,km09,bs10}. We find that in this case, the dissipation is used to create the required coherence in the atomic system.
The maximal stationary entanglement can be created on demand even for the resonant driving field by tuning the Rabi frequency of the field to the difference between the atomic transition frequencies. As a result, the atomic system evolves into a pure trapping state which is an asymmetric superposition of the degenerate energy states. The particular pure state into which the atomic system evolves depends upon the ratio of the damping rates of the atomic transitions and the detuning of the laser frequency from the atomic transition frequency. The trapping effect results in the complete population inversion between the dressed states of the system. In other words, the maximal steady state entanglement is generated when the population of the atomic system is trapped in a pure superposition state.

In the third scenario, we assume that the cavity modes are coupled to different atomic transitions. The new feature of this scenario is that now the generation of entanglement is independent of the population of the dressed states. The necessary condition for entanglement is the creation of coherence between the atomic transitions, the coherence that can be created by the SGC.

Finally, in the fourth scenario, we consider the most general configuration in which each of the cavity modes is coupled to both atomic transitions. We show that this scenario can be treated as a combination of the second and third scenarios, and find that the generation of entanglement depends now on the mutual polarization of the atomic dipole moments. Depending on whether the transition dipole moments are parallel or anti-parallel, the entanglement can be enhanced (reduced) by the constructive (destructive) interference between the atomic transition amplitudes.

The paper is organized as follows. We begin in Sec.~\ref{sec2} with a description of the proposed schemes for the generation of entanglement between two nondegenerate cavity modes and derive the master equation for the reduced density operator of the cavity modes. In Sec.~\ref{sec3}, we study the generation and enhancement of entanglement between the cavity modes for different coupling configurations of the cavity modes to the atomic transitions. We are particularly interested in the role of the mutual polarization of the atomic dipole moments and the conditions for the generation of a large stationary entanglement between the modes. The physical origin of entanglement between the cavity modes is explained in terms of population trapping in a linear superposition of the atomic levels. Finally, we summarize our results in Sec.~\ref{sec4}.

\section{General formalism}\label{sec2}

\begin{figure}[hbt]
\begin{center}
\includegraphics[width=7cm,keepaspectratio,clip]{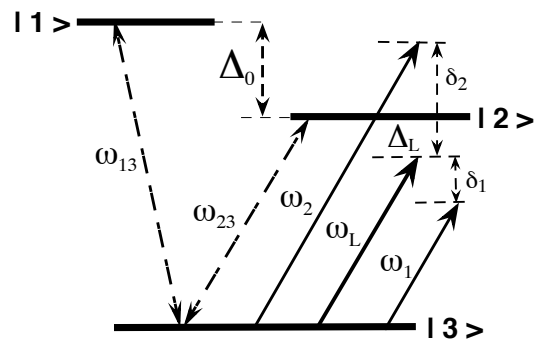}
\caption{Schematic diagram of the atomic levels and one of possible coupling configurations of the laser and the cavity fields. A laser field of frequency $\omega_{L}$ drives the $\ket 3 \rightarrow \ket 2$ transition with detuning~$\Delta_{L}$ and two non-degenerate cavity modes of frequencies~$\omega_{1}$ and $\omega_{2}$ couple to the driven transition with detunings $\delta_{1}$ and~$\delta_{2}$ from the laser frequency.}
\label{fig1a}
\end{center}
\end{figure}

We consider a three-level atom located inside a two-mode cavity. The atom is modelled as a V-type system with ground state $\left| 3 \right\rangle$, and two excited states $\left| 1 \right\rangle$ and $\left| 2\right\rangle$ separated in frequency by $\Delta_{0}=\omega_{13}-\omega_{23}$, where $\omega_{13}$ and $\omega_{23}$ are atomic transition frequencies between states $\ket 1 \leftrightarrow\ket3$ and $\ket 2 \leftrightarrow\ket3$, respectively. We shall assume that $\omega_{13}>\omega_{23}$ so that $\Delta_{0}$ is positive. This choice, of course, involves no loss of generality. The atom acts as a coupling (or entangling) medium that couples two non-degenerate cavity modes of frequencies~$\omega_{1}$ and $\omega_{2}$ through the interaction of the modes with the atomic dipole transitions $\left| 1\right\rangle \leftrightarrow \left| 3 \right\rangle $ and $\left| 2\right\rangle \leftrightarrow \left| 3 \right\rangle$. In addition, the transition $\left| 2\right\rangle \leftrightarrow \left| 3 \right\rangle$ is driven by a strong laser field of angular frequency $\omega_{L}$ and the amplitude determined by the Rabi frequency $2\Omega$, as illustrated in Fig.~\ref{fig1a}. The dipole moments of the two allowed atomic transitions can be orthogonal or non-orthogonal to each other. The latter case can lead to quantum interference effects induced by the SGC.
The cavity modes can simultaneously couple to one of the atomic transitions or to different transitions. One can also arrange a situation in which each of the cavity modes could couple to both of the atomic transitions. In this case, the coupling and the resulting entanglement between the modes can depend on whether the transition dipole moments are parallel or anti-parallel to each other.

For an open cavity in which the atom and the cavity modes are coupled to the outside vacuum modes, the dynamics of the driven atom plus the cavity modes is conveniently described by the density operator $\rho$, which in a frame rotating with the laser frequency frequency~$\omega_{L}$ satisfies the following master equation $(\hbar =1)$
\begin{eqnarray}
\frac{d}{dt}\rho = -i[H_{c} + H_{a} +V,\rho]+ L_{c}\rho + L_{a}\rho ,\label{e1}
\end{eqnarray}
where
\begin{equation}
H_{c} = -\delta_{1}a^{\dagger}_{1}a_{1} + \delta_{2}a^{\dagger}_{2}a_{2} \label{e2a}
\end{equation}
is the free Hamiltonian of the cavity modes,
\begin{equation}
H_{a} = \left( \Delta_{L} +\Delta_{0} \right)A_{11} +\Delta_{L} A_{22} -\Omega \left(A_{23} +A_{32}\right) \label{e2}
\end{equation}
is the Hamiltonian of the driven atom,
\begin{eqnarray}
V = \left(g_{1}a_{1} +g_{2}a_{2}\right)A_{23} + \left(g_{3}a_{1} +g_{4}a_{2}\right)A_{13} + {\rm H.c.} \label{e3}
\end{eqnarray}
is the interaction Hamiltonian of the cavity modes with the atomic transitions,
\begin{equation}
L_{c}\rho =\sum\limits_{j=1}^{2}\kappa_{j}\left(2a_{j}\rho a_{j}^{\dagger}-a_{j}^{\dagger }{{a}_{j}}\rho -\rho a_{j}^{\dagger }a_{j}\right)  \label{e4}
\end{equation}
and
\begin{eqnarray}
L_{a}\rho &=& \gamma_{1}\left[A_{31},\rho A_{13}\right] + \gamma_{2}\left[A_{32},\rho A_{23}\right] \nonumber\\
&+& \eta ([A_{31},\rho {A}_{23}]+[A_{32},\rho A_{13}]) + {\rm H.c.} \label{e5}
\end{eqnarray}
are operators representing the damping of the cavity-field modes by cavity decay with rates $\kappa_{1}$ and $\kappa_{2}$, and of the atomic transitions by spontaneous emission with rates $\gamma_{1}$ and $\gamma_{2}$. The parameters $g_{i}\, (i=1,2,3,4)$ are coupling strengths of the cavity modes to the atomic transitions. We assume that in general the modes couple with strengths $g_{1}$ and $g_{2}$ to the transition $\ket{2}\leftrightarrow \ket{3}$ and and also can be simultaneously coupled to the $\ket{1} \leftrightarrow \ket{3}$ transition with strengths $g_{3}$ and $g_{4}$, respectively.

The coefficient $\eta = p\sqrt{\gamma_{1}\gamma_{2}}$ is a measure of the amount of coherence, the so-called SGC, induced by dissipation between the $\ket{1} \leftrightarrow\ket{3}$ and $\ket{2} \leftrightarrow\ket{3}$ atomic transitions. The source of this coherence has an obvious interpretation. Namely, spontaneously emitted photon on one of the atomic transition drives the other transition. The degree of the coherence, measured by the coefficient $\eta$, depends explicitly on the mutual polarization of the transition dipole moments with $p = \cos\theta $, where $\theta$ is the angle between the two dipole moments. Thus, $p=0$ when the transition dipole moments are orthogonal to each other and~$p$ attains its maximal value of $p=\pm 1$ when the dipole moments are parallel or anti-parallel to each other. Obviously, the SGC vanishes when $p=0$ and attains maximal value when~$p=\pm 1$.

The parameter $\Delta_{L} =\omega_{23} -\omega_{L}$ is the detuning of the laser frequency $\omega_{L}$ from the atomic transition frequency $\omega_{23}$, $\delta_{1} = \omega_{L} -\omega_{1}$ and  $\delta_{2} = \omega_{2} -\omega_{L}$ are detunings of the cavity modes $\omega_{1}$ and $\omega_{2}$ from the laser frequency, respectively; $A_{ij} =\ket{i} \bra{j}$ are the atomic transition operators between energy states $\ket{i}$ and $\ket{j} ,\ (i,j =1,2,3)$ of the atom.

Since the transition $\left| 2\right\rangle \leftrightarrow \left| 3\right\rangle$ is driven by a strong, nearly resonant laser field, it is convenient to work in the dressed-state picture~\cite{ct77,ct92}. We introduce dressed states, which are the eigenstates of the Hamiltonian (\ref{e2}):
\begin{eqnarray}
\ket{\tilde 1} &=& \ket{1} ,\nonumber\\
\ket{\tilde 2} &=& \sin\phi \left| 2 \right\rangle - \cos\phi \left| 3 \right\rangle ,\nonumber\\
\ket{\tilde 3} &=& \cos\phi \left| 2 \right\rangle + \sin\phi \left| 3 \right\rangle ,\label{e6}
\end{eqnarray}
where
\begin{eqnarray}
\cos^{2}\phi =\frac{1}{2} + \frac{\Delta_{L}}{2\Omega_{0}} ,\label{e7}
\end{eqnarray}
and $\Omega_{0} =\sqrt{\Delta_{L}^{2} + 4\Omega^{2}}$ is the Rabi frequency of the detuned field. In the dressed-state basis, the operators $A_{ij}$ are replaced by dressed-state operators $R_{ij} =\left|\tilde{i}\right\rangle \left\langle \tilde{j} \right|$, and the density operator of the system can be transformed to the dressed-atom picture by the unitary transformation
\begin{eqnarray}
\tilde{\rho} =\exp\left(i\tilde{H}_{0}t\right)\rho\exp\left(-i\tilde{H}_{0}t\right) ,\label{e9}
\end{eqnarray}
where
\begin{eqnarray}
\tilde{H}_{0} = (\Delta_{L} +\Delta_{0})R_{11}+\Omega_{0}R_{z} - \delta_{1}a^{\dagger}_{1}a_{1}  + \delta_{2}a^{\dagger}_{2}a_{2} ,\label{e10}
\end{eqnarray}
and $R_{z} = (R_{22} -R_{33})/2$ is the population inversion operator between the dressed states $|\tilde{2}\rangle$ and  $|\tilde{3}\rangle$.

Applying the unitary transformation (\ref{e9}), we find that the commutator part of the master equation for $\tilde{\rho}$ contains explicitly time dependent terms that oscillate at frequencies $\delta_{1}$ and $\delta_{2}$, and the atomic dissipative part contains terms oscillating with $\Omega_{0}$ and $2\Omega_{0}$. In the limit of large Rabi frequency $\Omega_{0}\gg g_{i},\gamma_{i}$, the oscillating terms in the dissipative part make contributions of order $\gamma_{i}/\Omega_{0}$, where $i=1,2$. These terms can be neglected in the secular approximation. The errors of the secular approximation are of order $\gamma_{i}/\Omega_{0}$ and $g_{i}/\Omega_{0}$. Thus, it is reasonable to neglect these terms on time scales $t\gg \gamma_{i}^{-1}$ when $\Omega_{0}\gg g_{i},\gamma_{i}$. This approximation permits important mathematical simplifications, and "exact" solutions for the steady-state density matrix elements may be obtained that could provide immediate insight into the physics involved in the problem.

Thus, the maser equation in the dressed-atom basis and under the secular approximation simplifies to
\begin{eqnarray}
\frac{d}{dt}\tilde{\rho} = -i\!\left[\tilde{V},\tilde{\rho}\right] + L_{d}\tilde{\rho} + L_{c}\tilde{\rho} ,\label{e11}
\end{eqnarray}
where
\begin{eqnarray}
\tilde{V} &=& \left\{d_{1}\!\left[\sin (2\phi) R_{z}\!+\!\sin^{2}\!\phi R_{23}{\rm e}^{i\Omega_{0} t}\!-\! \cos^{2}\!\phi R_{32}{\rm e}^{-i\Omega_{0} t}\right]\right.\nonumber\\
&+&\left. d_{2}\left(\sin \phi R_{13}{\rm e}^{i[\Delta_{0}+\frac{1}{2}(\Omega_{0}+\Delta_{L})]t}\right.\right. \nonumber\\
&&\left.\left. -\cos \phi R_{12}{\rm e}^{i[\Delta_{0}-\frac{1}{2}(\Omega_{0}-\Delta_{L})]t}\right)\right\} + {\rm H.c.} \label{e12}
\end{eqnarray}
is the interaction of the dressed atom with the cavity modes with
\begin{eqnarray}
d_{1} &=& g_{1}a_{1}{\rm e}^{i\delta_{1} t} + g_{2}a_{2}{\rm e}^{-i\delta_{2} t} ,\nonumber\\
d_{2} &=& g_{3}a_{1}{\rm e}^{i\delta_{1} t} + g_{4}a_{2}{\rm e}^{-i\delta_{2} t} ,\label{e13}
\end{eqnarray}
and
\begin{align}
L_{d}\tilde{\rho} &=\gamma_{1}\!\left(\sin^{2}\!\phi [R_{31},\tilde{\rho} R_{13}]
+ \cos^{2}\!\phi [R_{21},\rho R_{12}] +{\rm H.c.}\right) \nonumber\\
&+\gamma_{2}\sin^{2}(2\phi)\left([R_{z},\tilde{\rho} R_{z}] +{\rm H.c.}\right) \nonumber\\
&+\gamma_{2}\!\left(\sin^{4}\!\phi \left[R_{32},\tilde{\rho} R_{23}\right]
+ \cos^{4}\!\phi [R_{23},\tilde{\rho} R_{32}] + {\rm H.c.}\right) \nonumber\\
&+\eta_{0}\sin^{2}\!\phi \left([R_{31},\tilde{\rho} R_{23}]+[R_{32},\tilde{\rho} R_{13}] +{\rm H.c.}\right) \nonumber\\
&+\eta_{0}\cos^{2}\!\phi\left([R_{21},\tilde{\rho} R_{22}]
+[R_{22},\tilde{\rho} R_{12}] + {\rm H.c.}\right)
\end{align}
is an operator representing the damping of the dressed-atom system.

Obviously, the cavity damping term remains unchanged under the dressed-atom transformation, but the atomic dynamics are now determined in terms of the dressed-atom operators.
Here, we are interested in the case of the two cavity modes being non-degenerated i.e., $\omega_{1}\neq \omega_{2}$, for which the time dependence of $\tilde{V}$ is quite complicated. This renders the master equation difficult to solve exactly, except in a special case of a weak coupling of the cavity modes to the atomic transitions, $g_{i}\ll \Omega_{0}$. In this case, we can treat the interaction as a weak perturbation to the strong atom-laser interaction and find, after tracing over the atomic variables, that the effective master equation for the reduced density operator of the cavity modes, $\rho_{c} = {\rm Tr}_{A}\tilde{\rho}$, is of the form
\begin{align}
\frac{d}{dt}\rho_{c} &= i\sum_{j=1}^{2}\left(\delta_{12} -\bar{B}_{j}\right)\left[a_{j}^{\dagger }a_{j},\rho_{c}\right] -i\sum_{j=1}^{2}\bar{A}_{j}\left[a_{j}a_{j}^{\dagger },\rho_{c}\right]\nonumber\\
&+ \sum_{j=1}^{2}\left(\tilde{B}_{j} +\kappa_{j}\right)\left(2a_{j}\rho_{c}a_{j}^{\dagger }
-a_{j}^{\dagger }a_{j}\rho_{c} -\rho_{c}a_{j}^{\dagger }a_{j}\right) \nonumber\\
&+ \sum_{j=1}^{2}\tilde{A}_{j}\left(2a_{j}^{\dagger }\rho_{c}a_{j} - \rho_{c}a_{j}a_{j}^{\dagger}
-a_{j}a_{j}^{\dagger }\rho_{c}\right) \nonumber\\
&+\sum_{j\ne j' = 1}^{2}\left\{C_{j}a_{j}^{\dagger }a_{j'}^{\dagger }\rho_{c}
+D_{j}\rho_{c}a_{j'}^{\dagger}a_{j}^{\dagger}\right. \nonumber\\
&\left. - \left(C_{j}+D_{j}\right)a_{j'}^{\dagger }\rho_{c}a_{j}^{\dagger } + {\rm H.c.}\right\} ,\label{e14}
\end{align}
where $\delta_{12} = (\delta_{2} - \delta_{1})/2$, $\tilde{A}_{j}, \tilde{B}_{j}$ and $\bar{A}_{j}, \bar{B}_{j}$ are the real and imaginary parts of complex coefficients $A_{j}, B_{j}$, respectively. The coefficients $\tilde{A}_{j}$ and $\tilde{B}_{j}$ have obvious interpretation as absorption and gain rates, whereas $\bar{A}_{j}$ and $\bar{B}_{j}$ are radiative shifts of the cavity mode frequencies. Correspondingly, the complex coefficients $C_{j}$ and $D_{j}$ determine terms representing desired correlations between the cavity modes. The expressions for the coefficients depend strongly on the coupling configuration of the cavity modes to the atomic transitions and also on a particular choice of other parameters. The explicit analytical forms of the coefficients for different coupling configurations of the cavity modes to the atoms will be given in Sec.~\ref{sec3}.

The master equation (\ref{e14}) is of a form characteristic for a system composed of two field modes coupled to a multi-mode squeezed vacuum~\cite{df04}. For this reason, to quantify entanglement between the modes, we shall use the Duan's criterion~\cite{DL00}, which relates entanglement to squeezing between the modes. If the cavity modes were initially in a vacuum state, which is an example of a Gaussian state, the state of the modes governed by Eq.~(\ref{e14}) will remain a two-mode Gaussian state for all times $t$. The quantum statistics properties of a two-mode Gaussian state are conveniently studied in terms of quadrature operators of the two cavity modes
\begin{eqnarray}
X_{l} &=& \frac{1}{\sqrt{2}}\left(a_{l}^{\dagger}{\rm e}^{i\theta_{l}} + a_{l}{\rm e}^{-i\theta_{l}}\right) ,\nonumber\\
Y_{l} &=& \frac{i}{\sqrt{2}}\left(a_{l}^{\dagger}{\rm e}^{i\theta_{l}} - a_{l}{\rm e}^{-i\theta_{l}}\right) ,\quad l=1,2, \label{e15}
\end{eqnarray}
where $\theta_{l}$ is the phase angles of the modes. If we introduce two operators
\begin{eqnarray}
u = aX_{1} -\frac{1}{a}X_{2} ,\quad v = aY_{1} +\frac{1}{a}Y_{2} ,\label{e16}
\end{eqnarray}
where $a$ is a state-dependent real number, then, according to the Duan's criterion, a two-mode Gaussian state is entangled if and only if the sum of the variances $\Sigma = \langle
(\Delta \hat{u})^{2}\rangle +\langle (\Delta \hat{v})^{2}\rangle$ satisfies the inequality
\begin{equation}
\Sigma = 2n a^{2} + 2m/a^{2} -4c < a^{2} +\frac{1}{a^{2}} ,\label{e17}
\end{equation}
with $a^{2} = \sqrt{(2m-1)/(2n-1)}$, $n = \langle a_{1}^{\dagger }a_{1}\rangle +1/2$, $m = \langle a_{2}^{\dagger }a_{2} \rangle +1/2$, and $c = |\langle a_{1}a_{2}\rangle|$.
Since the right-hand side of Eq.~(\ref{e17}) is a positive number, we may introduce a parameter
\begin{equation}
\Upsilon =\Sigma -a^{2} -\frac{1}{a^{2}} ,\label{e18}
\end{equation}
and then the condition for entanglement between the cavity modes is that the parameter $\Upsilon$ must be negative.

From Eqs.~(\ref{e17}) and (\ref{e18}) it is obvious that in order to calculate the parameter $\Upsilon$, it is necessary to have available the cavity field correlation functions $n, m$ and $c$. These correlation functions are readily found using the master equation~(\ref{e14}), from which we can derive equations of motion for the required correlation functions and find that they satisfy a set of coupled differential equations
\begin{align}
\frac{d}{dt}\langle a_{j}^{\dagger }{a}_{j} \rangle &= -\left(\Gamma_{j}+\Gamma_{j}^{*}\right)\langle a_{j}^{\dagger }a_{j} \rangle \nonumber\\
&+ \chi_{j}\langle a_{1}^{\dagger }a_{2}^{\dagger } \rangle + \chi_{j}^{\ast}\langle a_{1}a_{2}\rangle + 2\tilde{A}_{j} ,\nonumber\\
\frac{d}{dt}\langle a_{1}a_{2}\rangle &= - \left(\Gamma_{1} + \Gamma_{2}\right)\langle a_{1}a_{2}\rangle + \chi_{2}\langle a_{1}^{\dagger }a_{1}\rangle \nonumber\\
&+ \chi_{1}\langle a_{2}^{\dagger }a_{2}\rangle + \left(C_{1} + C_{2}\right) ,\label{e18a}
\end{align}
where $\Gamma_{j} = \kappa_{j} + i\delta_{12} -(A_{j} - B_{j})$ and $\chi_{j} = C_{j} - D_{j}$.
The set of the differential equations (\ref{e18a}) can be easily solved for arbitrary initial conditions. Since we are interested in a stationary entanglement between the cavity modes, we analyze the stability condition and find that the system is stable and reaches its steady-state as $t\rightarrow \infty$ when
\begin{eqnarray}
{\rm Re}\left[\Gamma_{1} +\Gamma_{2} - \sqrt{\left(\Gamma_{1} - \Gamma_{2}^{*}\right)^{2}+4\chi_{1}\chi_{2}^{*}}\right] > 0 .\label{e18b}
\end{eqnarray}
The above stability condition may be simplified substantially for particular choices of the detunings and the Rabi frequency such as $\delta_{1},\delta_{2}\gg\gamma_{i}$ and $\Omega_{0}\gg\gamma_{i}$. 
\section{Entanglement between cavity modes}\label{sec3}

It is clear from Eq.~(\ref{e14}) that the dynamics and entanglement of the cavity modes are a sensitive function of the properties of the driven atomic system. In order to study this dependence, we shall examine four scenarios of the coupling configuration of the cavity modes to the atomic transitions, two scenarios in which both modes couple to the same driven atomic transition and the other two in which the cavity modes are coupled to different transitions. A particular attention will be paid to the role of a specific driving of the atoms and the SGC in entangling the cavity modes.

\subsection{The case of both modes coupled to the driven transition}\label{sec3a}

In this section, we examine the entanglement properties of the cavity modes when both modes are coupled to only one of the atomic transitions, the laser driven transition $\left| 2 \right\rangle \leftrightarrow \left| 3 \right\rangle$, as illustrated in Fig.~\ref{fig1a}. In other words, all the fields couple to only one of the atomic transition. This is achieved by putting the coupling strengths $g_{3}$ and~$g_{4}$ in the Hamiltonian~(\ref{e3}) equal to zero. We shall be particularly interested in the generation of entanglement between the cavity modes when the coupling system is reduced to a simple two-level system and the role of the spontaneous emission in coupling of the two-level system to the auxiliary level $\ket 1$. Therefore, we consider separately two cases of orthogonal $(p=0)$ and non-orthogonal $(p\neq 0)$ dipole moments of the atomic transitions. When the dipole moments are orthogonal to each other, $p=0$, and then the atomic transition $\left| 1\right\rangle \leftrightarrow \left| 3 \right\rangle$ decouples from the driven transition. In this case, the system reduces to that of a driven two-level atom. On the other hand, when the dipole moments are nonorthogonal, $p\neq 0$, and then the spontaneous emission on the $\ket 1\leftrightarrow \ket 3$ can influence on the two-level dynamics of the driven $\ket 2\leftrightarrow \ket 3$ transition.

We start by introducing the explicit form of the coefficients of the master equation (\ref{e14}), which read
\begin{align}
A_{1} &= g_{1}^{2}\left[-\frac{1}{4}F_{1}(\delta_{1})\sin 2\phi +\frac{ f_{1}^{*}(-\delta_{1})\rho_{33}^{s}}{f_{12}^{*}(-\delta_{1})-\eta _{0}^{2}}\cos^{4}\phi\right.\nonumber\\
&\left. +\frac{f_{1}(\delta_{1})\rho_{22}^{s} - \eta_{0}\rho _{12}^{s}}{f_{12}(\delta_{1}) -\eta _{0}^{2}}\sin^{4}\phi \right] ,\nonumber\\
B_{1} &= g_{1}^{2}\left[ -\frac{1}{4}F_{2}(\delta_{1})\sin 2\phi +\frac{f_{1}(\delta_{1})\rho_{33}^{s} }{f_{12}(\delta_{1})-\eta _{0}^{2}}\sin^{4}\phi\right.\nonumber\\
&\left. + \frac{f_{1}^{*}(-\delta_{1})\rho_{22}^{s} -\eta_{0}\rho_{21}^{s}}{f_{12}^{*}(-\delta_{1})-\eta _{0}^{2}}\cos^{4}\phi\right] ,\nonumber\\
C_{1} &= \frac{1}{4}g_{1}g_{2}\sin 2\phi \left[F_{2}(\delta_{2})+\frac{f_{1}(\delta_{2})\rho_{33}^{s}}{f_{12}(\delta_{2})-\eta _{0}^{2}}\right. \nonumber\\
&\left. +\frac{f_{1}^{*}(-\delta_{2})\rho_{22}^{s} -\eta_{0}\rho_{21}^{s}}{f_{12}^{*}(-\delta_{2})-\eta_{0}^{2}} \right] ,\nonumber\\
D_{1} &= \frac{1}{4}g_{1}g_{2}\sin 2\phi \left[F_{1}(\delta_{2})
+\frac{f_{1}^{\ast}(-\delta_{2})\rho_{33}^{s}}{f_{12}^{*}(-\delta_{2})-\eta _{0}^{2}}\right. \nonumber\\
&\left. +\frac{f_{1}(\delta_{2})\rho_{22}^{s} -\eta_{0}\rho_{12}^{s}}{f_{12}(\delta_{2}) -\eta _{0}^{2}}\right] ,\label{e19}
\end{align}
where
\begin{align}
F_{1}(\delta_{j}) &= \left[M_{32}(\delta_{j})\!-\!M_{22}(\delta_{j})\right]\!\rho_{22}^{s}
-\!\left[M_{33}(\delta_{j})\!-\!M_{23}(\delta_{j})\right]\!\rho_{33}^{s} \nonumber\\
&+\left[M_{34}(\delta_{j}) -M_{24}(\delta_{j})\right]\!\rho_{12}^{s} ,\nonumber\\
F_{2}(\delta_{j}) &= \left[M_{32}(\delta_{j})\!-\!M_{22}(\delta_{j})\right]\!\rho_{22}^{s}
-\!\left[M_{33}(\delta_{j})\!-\!M_{23}(\delta_{j})\right]\!\rho_{33}^{s} \nonumber\\
&+\left[M_{35}(\delta_{j}) -M_{25}(\delta_{j})\right]\!\rho_{21}^{s} ,
\end{align}
and
\begin{align}
f_{12}(\pm\delta_{j}) = f_{1}(\pm\delta_{j})f_{2}(\pm\delta_{j}) ,\quad j=1,2,
\end{align}
with
\begin{align}
f_{1}(\pm\delta_{j}) &= \gamma_{1} +\gamma_{2}\cos^{2}\!\phi + i\!\left(\Delta_{0} +\frac{1}{2}\left(\Delta_{L} +\Omega_{0}\right) \pm\delta_{j}\right) ,\nonumber\\
f_{2}(\pm\delta_{j}) &= \gamma_{2}\left(1 +\frac{1}{2}\sin^{2}2\phi\right) + i\left(\Omega_{0} \pm\delta_{j}\right)  .\label{e20}
\end{align}
Here, $\rho_{22}^{s}$, $\rho_{33}^{s}$, $\rho_{12}^{s}$ are the steady-state values of the atomic density matrix elements under the condition of ignoring the effect of the weak coupling between the cavity modes and the atom, and $M_{mn}(\delta_{j})$ are elements of the inverse matrix of $U(\delta_{j})$:
\begin{widetext}
\begin{equation}
U(\delta_{j}) ={{\left( \begin{matrix}
  2\gamma_{1} +i\delta_{j}  & 0 & 0 & \eta_{0} & \eta_{0}  \\
 -2\gamma_{1}\cos^{2}\phi  & 2\gamma_{2}\sin^{4}\phi +i\delta_{j} & -2\gamma_{2}\cos^{4}\phi  & -\eta_{0}\cos 2\phi  & -\eta_{0}\cos 2\phi   \\
   -2\gamma_{1}\sin^{2}\phi  & -2\gamma_{2}\sin^{4}\phi  & 2\gamma_{2}\cos^{4}\phi +i\delta_{j} & -2\eta_{0}\sin^{2}\phi  & -2\eta_{0}\sin^{2}\phi   \\
   \eta_{0} & \eta_{0} & 0 & b+i\delta_{j} & 0  \\
   \eta_{0} & \eta_{0} & 0 & 0 & b^{*}+i\delta_{j}  \\
\end{matrix} \right)}} .
\end{equation}
\end{widetext}
where $b=\gamma_{1}+\gamma_{2}\sin^{2}\phi+i[\Delta_{0}-(\Omega_{0}-\Delta_{L})/2]$.

The remaining coefficients $A_{2}, B_{2}, C_{2}$ and $D_{2}$ are obtained from Eq.~(\ref{e19})  by exchanging $\delta_{1}$ with $-\delta_{2}$ and $g_{1}$ with $g_{2}$. We should point out here that in the derivation of the coefficients~(\ref{e19}), we have assumed that the states $\ket{\tilde 1}$ and $\ket{\tilde 3}$ are separated in energy by $\Delta_{0}+(\Omega_{0}+\Delta_{L})/2$, while the states $\ket{\tilde 1}$ and $\ket{\tilde 2}$ are separated in energy by $\Delta_{0}-(\Omega_{0}-\Delta_{L})/2$. Thus, in general, the dressed states are non-degenerate. However, by varying the Rabi frequency $\Omega_{0}$ or the splitting $\Delta_{0}$, one may turn the states $\ket{\tilde 1}$ and $\ket{\tilde 2}$ into degeneracy, whereas the states $\ket{\tilde 1}$ and $\ket{\tilde 3}$ will always remain far from resonance. This would happen when $\Delta_{0}= (\Omega_{0}-\Delta_{L})/2$. As we shall demonstrate in this paper, the degeneracy condition is an optimal condition for entanglement between the cavity modes.

Having defined the coefficients of the master equation for the case of both cavity modes coupled to the driven atomic transition, we now turn our attention to the possibility of generating a stationary entanglement between the modes. In doing that we shall consider separately two cases, $p=0$ and $p\neq 0$.

\subsubsection{The case of $p=0$}

\begin{figure}[hbt]
\includegraphics[width=\columnwidth,keepaspectratio,clip]{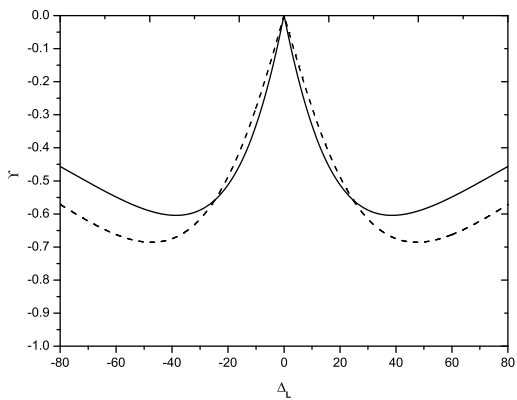}
\caption{The degree of entanglement $\Upsilon$ plotted as a function of $\Delta_{L}$ for the case corresponding to a two-level system, $g_{3}=g_{4}=0$ and $p=0$, with $\gamma_{2} = 0.02, \Omega = 50, \delta_{1} \approx \delta_{2} = 50, \kappa_{1} = \kappa_{2} = 0.63, g_{1} = g_{2} = 10$ and different $\delta_{12}$: $\delta_{12}=0$ (solid line), $\delta_{12}=-0.61$ (dashed line). All parameters are normalized to $\gamma_{1}$.}
\label{fig2}
\end{figure}

Let us first determine how much entanglement can be generated when the atom behaves as a two-level system. The master equation~(\ref{e14}) can be applied to this simplified case by putting $p=0$. Figure~\ref{fig2} shows the entanglement measure $\Upsilon$ as a function of $\Delta_{L}$ for $\eta_{0}=0$, fixed detunings $\delta_{1}, \delta_{2}$ and the Rabi frequency~$\Omega_{0}$. The figure shows that under resonant excitation, the cavity modes are separable and become entangled for an off-resonant excitation. The entanglement exhibits an interesting behavior, in that it has two maxima which occur for certain nonzero values of $\Delta_{L}$, and then rapidly declines thereafter. A small difference $\delta_{12}=-0.61$ between the detunings $\delta_{1}$ and $\delta_{2}$ is introduced to cancel the effect of the Stark shifts $\bar{A}_{j}$ and $\bar{B}_{j}$. As we see from the figure, the Stark shifts have a distractive effect on entanglement.

We would like to point out that the magnitude of the entanglement is not large and there are no parameter values at which the entanglement could reach the optimal value $\Upsilon =-1$. Moreover, the maximal entanglement occurs at large detunings, $\Delta_{L}\approx\pm 40\gamma_{1}$, at which the driving field is weakly coupled to the atoms. We shall demonstrate in the second scenario, that the magnitude can be enhanced to its optimal value $\Upsilon =-1$ by coupling the two-level system to the third level. To summarize, we briefly discuss the parameters characterizing the system and the ranges of these parameters experimentally accessible. The parameters are expressed in units of the spontaneous emission rate $\gamma$. In the case of alkali atoms, $\gamma$ is of the order of $10$ MHz. Driving lasers used in experiments are usually tunable, providing for arbitrary detuning $\Delta_{L}$, so that the range $\Delta_{L}\leq 100\gamma$ is easily accessible. The lasers are sufficiently powerful to generate Rabi frequencies up to $100\gamma$.

\subsubsection{The case of $p\neq 0$}\label{sec3b}

We now turn to illustrate the role of the SGC on entanglement creation between the cavity modes. We assume that the driven transition to which the cavity modes are coupled, is coupled by spontaneous emission to the auxiliary level $\ket{1}$. This coupling can occur for the case of non-orthogonal $(p\neq 0)$ dipole moments of the atomic transitions, and then the spontaneous emission on the $\ket{1}\leftrightarrow \ket{3}$ can influence on the two-level dynamics of the driven $\ket{2}\leftrightarrow \ket{3}$ transition.

\begin{figure}[hbt]
\includegraphics[width=\columnwidth,keepaspectratio,clip]{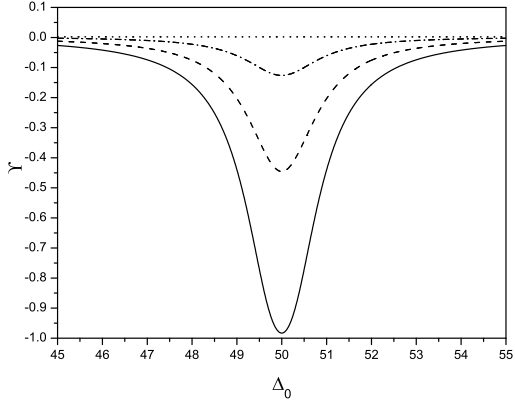}
\caption{The degree of entanglement $\Upsilon$ plotted as a function of $\Delta_{0}$ for $\Delta_{L} = 0$, $\gamma_{2} = 0.02$, $\Omega = 50$, $\delta_{1} \approx \delta_{2} = 50$, $\delta_{12} = -0.61$, $\kappa_{1} = \kappa_{2} = 0.63$, $g_{1} = g_{2} = 10$, and various values of $p$: $p=0.98$ (solid line), $p=0.7$ (dashed line), $p=0.4$ (dashed-dotted line), $p=0$ (dotted line). All parameters are normalized to $\gamma_{1}$.}
\label{fig3}
\end{figure}

Since the spontaneous emission on the atomic transitions occurs at different frequencies and with different rates, the created entanglement between the cavity modes may depend strongly on the splitting $\Delta_{0}$. As we shall see, the crucial for entanglement between the cavity modes is the relation between $\Omega_{0}$ and $\Delta_{0}$. Figure~\ref{fig3} illustrates the variation of $\Upsilon$ with gradually increasing $\Delta_{0}$ for the case of resonant driving, $\Delta_L=0$. We see that the cavity modes become entangled {\it only} for $p\neq 0$ and for a certain value of $\Delta_{0}=\Omega_{0}/2$, the entanglement becomes optimal. In terms of the energies of the dressed states, the condition of $\Delta_{0}=\Omega_{0}/2$ corresponds to the situation where the dressed states $\ket{\tilde 1}$ becomes degenerate with the dressed state $\ket{\tilde 2}$~\cite{SM99,FZ01}. The condition of $p\neq 0$ corresponds to the presence of direct coupling between the states $\ket{\tilde 1}$ and $\ket{\tilde 2}$. Note that this coupling is induced by the dissipative process of spontaneous emission. Since this is a resonant coupling, it creates a strong coherence between the states $\ket{\tilde 1}$ and $\ket{\tilde 2}$. Under this circumstance, the modes become strongly entangled and the degree of entanglement is maximal in comparison with Fig.2. The amount of the generated entanglement depends also on the ratio of the spontaneous emission rates, $\gamma_{2}/\gamma_{1}$, and the maximal entanglement of $\Upsilon \approx -1$ is achieved at $\Delta_{0}=\Omega_{0}/2$ and $p\approx 1$ for $\gamma_{2}\ll\gamma_{1}$. In other words, a large entanglement occurs when the most of the population resides in the driven transition rather than in the undriven transition. We may summarize that by using carefully designed driving, such that $\Delta_{0}=\Omega_{0}/2$ and carefully chosen atoms, such that $\gamma_{2}\ll\gamma_{1}$, a large entanglement can be produced between the cavity modes via dissipation created coherence in the atoms.

We now proceed to explain the physical origin of the process responsible for entanglement of the cavity modes predicted in the above two scenarios. As we shall see, the physics of the process can be quantitatively explained on the level of the stationary population of the atomic system. In the first instance, a simple analytical expression can be derived for the master equation as follows. When the frequency difference $\delta$ and the Rabi frequency $\Omega_{0}$ are much larger than the damping rates of the atomic transitions, $\delta_{1}\approx \delta_{2} =\delta\gg\gamma_{i}$ and $\Omega_{0}\gg\gamma_{i}$, the real parts of the parameters (\ref{e19}) become negligible, i.e. $\tilde{A}_{j}=\tilde{B}_{j}=\tilde{C}_{j}=\tilde{D}_{j}\approx 0$, and the imaginary parts become $\bar{A}_{j}\approx \bar{B}_{j}$ and $\bar{C}_{j}=-\bar{D}_{j}$. It is then straightforward to show that the master equation~(\ref{e14}) may be approximated~by
\begin{align}
\frac{d}{dt} \rho_{c} =& -i\left(\delta_{12} +2\bar{A}\right)\left[a_{1}^{\dagger }a_{1} +a_{2}^{\dagger}a_{2}, \rho_{c}\right] \nonumber\\
&-i\bar{D}\left[a_{1}^{\dagger }a_{2}^{\dagger} + a_{1}a_{2}, \rho_{c}\right] + L_{c}\rho_{c} ,\label{e22}
\end{align}
where
\begin{eqnarray}
\bar{A} &=& \frac{g^{2}\Omega_{0}\left(1+\cos^{2}2\phi\right)}{4\left(\Omega _{0}^{2}-\delta^{2}\right)}\left( \rho _{22}^{s}-\rho _{33}^{s}\right) ,\nonumber\\
\bar{D} &=& \frac{g^{2}\Omega_{0}\sin^{2}2\phi}{2\left(\Omega _{0}^{2}-{{\delta }^{2}}\right)}( \rho _{22}^{s}-\rho _{33}^{s}) ,\label{e23}
\end{eqnarray}
and, for simplicity, we have assumed equal coupling constants $g_{1}=g_{2}=g$.

This shows that the atomic variables contribute to the coherent evolution of the cavity modes and the only relaxation in the system is the damping of the cavity modes.
A choice of $\delta_{12} =-2\bar{A}$ simplifies further the master equation and leaves only the parametric amplifying term in its commutator part. This term is responsible for correlations and so for entanglement between the modes. The magnitude of entanglement attains maximal value when $\bar{D}$ maximizes. It is evident from Eq.~(\ref{e23}) that the parameter $\bar{D}$ is different from zero only if the population is unequally distributed between the dressed states.
Thus, the only one factor determines the magnitude of entanglement between the cavity mode, {\it the population must be inverted between the dressed states} of the system. For the case of $p=0$, this can be achieved if the laser frequency is detuned from the atomic transition frequency $\omega_{23}$. It is interesting that the entanglement is created without any coherence between the dressed states. There is no coherence between the dressed states since the Rabi frequency $\Omega_{0}$ is much larger than all relaxation rates, $\Omega_{0}\gg\gamma_{i},\kappa_{i}$. However, we should point out that in the case of an off-resonat driving, there is a coherence between the bare atomic states. Thus, one can argue that the predicted entanglement actually occurs due to a non-zero coherence between the bare atomic states.

To calculate the population inversion between the dressed states, we introduce density matrix elements with respect to the three atomic dressed states in the absence of the cavity modes, denoting $\bra{\tilde 1}\tilde{\rho}\ket{\tilde 2}$ by $\rho_{12}$, etc. The equations of motion are 
\begin{align}
\dot{\rho}_{11} &=-2\gamma_{1}\rho_{11} - \eta_{0}(\rho_{12} + \rho_{21}) ,\nonumber\\
\dot{\rho}_{22} &= 2\gamma_{1}\cos^{2}\phi \rho_{11} +2\gamma_{2}\left(\cos^{4}\phi \rho_{33}-\sin^{4}\phi\rho_{22}\right) \nonumber\\
&+\eta_{0}\cos 2\phi (\rho_{12}+\rho_{21}),\nonumber\\
\dot{\rho}_{33} &= 2\gamma_{1}\sin^{2}\phi\rho_{11}-2\gamma_{2}\left(\cos^{4}\phi \rho_{33}
-\sin^{4}\phi \rho_{22}\right) \nonumber\\
&+2\eta_{0}\sin^{2}\phi (\rho_{12}+\rho_{21}) ,\nonumber\\
\dot{\rho}_{12} &=-\left\{\gamma_{1}+\gamma_{2}\sin^{2}\phi +i\left[\Delta_{0} -\frac{1}{2}\left(\Omega_{0}-\Delta_{L}\right)\right]\right\}\!\rho_{12} \nonumber\\
&-\eta_{0}(\rho_{11}+\rho_{22}) .\label{e24}
\end{align}
It is evident from the above equations that the coherence~$\rho_{12}$ induced by spontaneous emission oscillates with frequency $\Delta_{0} -(\Omega_{0}-\Delta_{L})/2$. This fact has the obvious physical meaning that the coherence attains maximal value when $\Delta_{0} -(\Omega_{0}-\Delta_{L})/2 =0$. For $\Delta_{L}=0$, the coherence maximizes at $\Delta_{0}=\Omega_{0}/2$ and simultaneously the factor $\sin^2\phi$ in the coefficient $\bar{D}$ equals to 1, consequently the value at which the entanglement, shown in Fig.~\ref{fig3}, attains the maximal value.

In the steady-state, the dressed state population difference can be worked out explicitly for both $p=0$ and $p\neq 0$. For the case of $p=0$, the steady state population difference is given by the expression
\begin{eqnarray}
\rho_{22}^{s}-\rho _{33}^{s} = \frac{\cos^{4}\phi - \sin^{4}\phi}{\cos^{4}\phi + \sin^{4}\phi} ,\label{e25}
\end{eqnarray}
which clearly shows that the populations among the dressed states are imbalanced only for a nonzero detuning $\Delta_{L}\neq 0 \, (\phi\neq \pi/4)$. In this case the parameter $\bar{D}$ responsible for the nonlinear coupling between the modes is different from zero. It is easy to check that the maximal entanglement seen in Fig.~\ref{fig2} is attained at the detunings corresponding to the maximal value of $\bar{D}$. Thus, we have a simple physical interpretation of the entanglement creation by a detuned laser field.

We stress that in the case of the detuned driving $(\Delta_{L}\neq 0)$ and in the limit $p=0$, i.e. in the two-level situation, the population is unequally distributed between the dressed states, but it is not possible to produce atoms in a pure dressed state in which $| \rho_{22}^{s}-\rho_{33}^{s}|=1$ and at the same moment having the coefficient $\bar{D}$ different from zero.
However, for the case of thee-level atoms with $p=1$, it is possible to have $| \rho_{22}^{s}-\rho_{33}^{s}|=1$, in which case the population is trapped in one of the dressed states. The condition of the population trapping is unique to the SGC and can be achieved even for a resonant driving, $\Delta_{L}=0$.

We now proceed to evaluate the population inversion when $p=1$. A careful analysis of the steady-state solution shows that in the case of the level crossing at $\Delta_{0}=\Omega_{0}/2$ and in the limit $p=1$, the population is not trapped in one of the dressed states but rather in one of linear superpositions
\begin{eqnarray}
\ket s &=& \alpha\ket{\tilde 2} +\beta\ket{\tilde 1} ,\nonumber\\
\ket a &=& \beta\ket{\tilde 2} -\alpha\ket{\tilde 1} ,\label{e26}
\end{eqnarray}
where
\begin{eqnarray}
\alpha = \left(\frac{\gamma_{2}\sin^{2}\phi}{\gamma_{1} +\gamma_{2}\sin^{2}\phi}\right)^{\frac{1}{2}} ,\quad \beta = \left(\frac{\gamma_{1}}{\gamma_{1} +\gamma_{2}\sin^{2}\phi}\right)^{\frac{1}{2}} .\label{e27}
\end{eqnarray}

It is easy to check that at the level crossing condition and in the limit $p=1$, the population is trapped in the antisymmetric state~$\ket a$, i.e. $\rho_{aa}^{s}=1$ irrespective of the detuning $\Delta_{L}$ and the ratio between the damping rates $\gamma_{1}$ and $\gamma_{2}$. This result implies that the SGC is essential for the atomic system to be capable of achieving a pure state. In other words, the trapping effect is a direct manifestation of the presence of the SGC that can be employed to maintain the complete inversion between the dressed states even in the case of zero detuning between the laser and the atomic transition frequencies. If we incorporate the solution $\rho_{aa}^{s}=1$ into Eq.~(\ref{e23}), we find that the resulting coefficient $\bar{D}$ takes the form
\begin{eqnarray}
\bar{D} =  \frac{g^{2}\Omega_{0}}{2\left(\Omega _{0}^{2}-{{\delta }^{2}}\right)}\frac{\gamma_{1}\sin^{2}2\phi}{\gamma_{1}+\gamma_{2}\sin^{2}\phi} ,\label{e28}
\end{eqnarray}
from which one can easily show that the coefficient $\bar{D}$ is greatest when $\phi=\pi/4 \, (\Delta_{L}=0)$ and $\gamma_{2}\ll\gamma_{1}$.  This prediction clearly explains our numerical results presented in Fig.~\ref{fig3}.

To clarify the issue of the mechanism responsible for creation of the stationary entanglement between the cavity modes, we may refer to the equations of motion for the correlation functions (\ref{e18a}). It is straightforward to show that the limit of $\delta\gg\gamma_{i}$ and $\Omega_{0}\gg\gamma_{i}$, the only damping mechanism of the correlation functions is the cavity damping. Thus, the SGC facilities correlations between the cavity modes that then decay with the cavity damping to a stationary entangled state.

\subsection{The case of the modes coupled to different atomic transitions}\label{sec3c}

\begin{figure}[hbt]
\begin{center}
\includegraphics[width=7cm,keepaspectratio,clip]{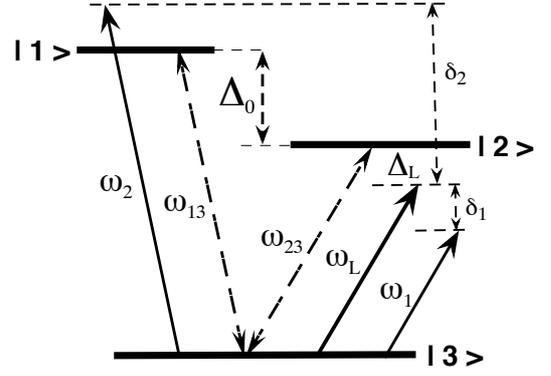}
\caption{Schematic diagram of the coupling configuration of the cavity modes and the driven laser field. The cavity mode of frequency~$\omega_{1}$ is coupled to the laser driven transition with detuning $\delta_{1}$ from the laser frequency, while the cavity mode of frequency $\omega_{2}$ is coupled to the undriven transition with detuning $\delta_{2}$ from the laser frequency.}
\label{fig1b}
\end{center}
\end{figure}

We now proceed to evaluate entanglement between the cavity modes when one of the cavity modes, $a_{1}$, is coupled to the driven $\ket 2 \leftrightarrow \ket 3$ transition and the other mode $a_{2}$ is coupled to the undriven transition $\ket 1 \leftrightarrow \ket 3$, as illustrated in Fig.~\ref{fig1b}.
In this case, the coupling strengths $g_{2}=g_{3}=0$, then the coefficients of the master equation (\ref{e14}) are of the form
\begin{align}
A_{1} &=g_{1}^{2}\left[-\frac{1}{4}F_{1}(\delta_{1})\sin 2\phi +\frac{\rho_{33}^{s}\cos^{4}\phi }{f_{2}^{*}(-\delta_{1})-\eta _{0}^{2}}\right.\nonumber\\
&\left.+\frac{\rho _{22}^{s}\sin^{4}\phi }{f_{2}^{*}(\delta_{1})-\eta _{0}^{2}} -\frac{\eta_{0}\rho_{12}^{s}\sin^{4}\phi }{f_{12}(\delta_{1})-\eta _{0}^{2}}\right] ,\nonumber\\
B_{1} &=g_{1}^{2}\left[ -\frac{1}{4}F_{2}(\delta_{1})\sin 2\phi +\frac{\rho_{33}^{s}\sin^{4}\phi }{f_{2}(\delta_{1})-\eta _{0}^{2}}\right.\nonumber\\
&\left. +\frac{f_{1}^{*}(-\delta_{1})\rho_{22}^{s} - \eta_{0}\rho _{21}^{s}}{f_{12}^{*}(-\delta_{1})-\eta _{0}^{2}}\cos^{4}\phi \right] ,\nonumber\\
C_{1} &= g_{1}g_{4}\sin \phi\cos^{2}\phi\left[F_{3}(\delta_{2})+\frac{f_{1}^{*}(-\delta_{2})\rho_{12}^{s} - \eta_{0}\rho_{11}^{s}}{f_{12}^{*}(-\delta_{2})-\eta _{0}^{2}}\right] ,\nonumber\\
D_{1} &= g_{1}g_{4}\sin \phi\cos^{2}\phi \left[F_{4}(\delta_{2})-\frac{\eta_{0}\rho_{33}^{s}}{f_{12}^{*}(-\delta_{2}) -\eta _{0}^{2}}\right] ,\label{e30}
\end{align}
with $F_{1}(\delta_{1})$ and $F_{2}(\delta_{1})$ given in Eq.~(\ref{e20}),
\begin{align}
F_{3}\left( \delta_{2} \right) &= \left[M_{32}(\delta_{2}) - M_{22}(\delta_{2})\right]\rho_{12}^{s} \nonumber\\
&+ \left[M_{35}(\delta_{2}) - M_{25}(\delta_{2})\right]\rho_{11}^{s} ,\nonumber\\
F_{4}\left( \delta_{2} \right) &=  \left[M_{31}(\delta_{2}) - M_{21}(\delta_{2})\right]\rho_{12}^{s} \nonumber\\
&+ \left[M_{35}(\delta_{2}) - M_{25}(\delta_{2})\right]\rho_{22}^{s} ,\label{e31}
\end{align}
and
\begin{align}
A_{2} &=g_{4}^{2}\left[h_{1}(\delta_{2}) +\frac{f_{2}(-\delta_{2})\rho_{11}^{s} -\eta _{0}\rho _{21}^{s}}{f_{12}(-\delta_{2})-\eta _{0}^{2}} \sin^{2}\phi \right] ,\nonumber\\
B_{2} &=g_{4}^{2}\left[h_{2}(\delta_{2}) +\frac{f_{2}(-\delta_{2})\rho_{33}^{s}\sin^{2}\phi}{f_{12}(-\delta_{2})-\eta _{0}^{2}}\right] ,\nonumber\\
C_{2} &= g_{1}g_{4}\sin \phi \cos^{2}\phi \left[ h_{3}(\delta_{1} )-\frac{\eta_{0}\rho _{33}^{s}}{f_{12}(-\delta_{1})-\eta _{0}^{2}} \right] ,\nonumber\\
D_{2} &= g_{1}g_{4}\sin \phi\cos^{2}\phi \left[h_{4}(\delta_{1})+\frac{f_{2}(-\delta_{1})\rho_{12}^{s}}{f_{12}(-\delta_{1})-\eta_{0}^{2}}\right] ,\label{e31a}
\end{align}
with
\begin{align}
h_{1}(\delta_{2}) &= \left[M_{42}\left(-\delta_{2} \right)\!\rho_{21}^{s}
+ M_{44}\left( -\delta_{2} \right)\!\rho_{11}^{s}\right]\cos^{2}\phi ,\nonumber\\
h_{2}(\delta_{2}) &= \left[M_{41}\left( -\delta_{2} \right)\!\rho_{21}^{s} + M_{44}\left( -\delta_{2} \right)\!\rho_{22}^{s}\right]\cos^{2}\phi ,\nonumber\\
h_{3}(\delta_{1}) &= M_{43}\left( -\delta_{1} \right)\!\rho_{33}^{s}
- M_{42}\left( -\delta_{1} \right)\!\rho_{22}^{s} - M_{45}\left( -\delta_{1} \right)\!\rho_{11}^{s} ,\nonumber\\
h_{4}(\delta_{1}) &= M_{43}\left( -\delta_{1} \right)\!\rho_{33}^{s}
- M_{42}\left( -\delta_{1} \right)\!\rho_{22}^{s}- M_{44}\left( -\delta_{1} \right)\!\rho_{12}^{s} .\label{e31b}
\end{align}

\begin{figure}[hbt]
\includegraphics[width=\columnwidth,keepaspectratio,clip]{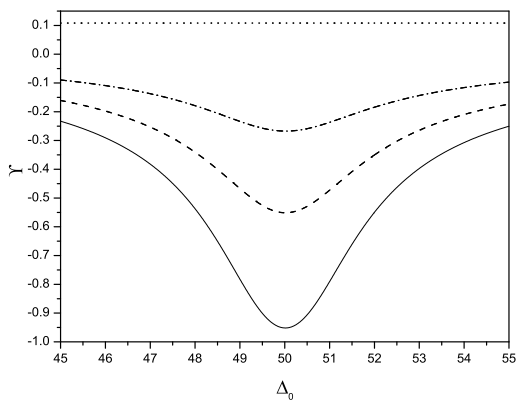}
\caption{The degree of entanglement $\Upsilon$ as a function of $\Delta_{0}$ for the case of the cavity modes coupled to different atomic transitions, $g_{2}=g_{3}=0$ and $g_{1} = g_{4} = 10$, with $\Delta_{L} =0, \gamma_{2}=2, \Omega = 50, \delta_{1} \approx \delta_{2} = 50, \delta_{12} =-0.38, \kappa_{1} = \kappa_{2} = 0.67$, and different $p$: $p=0.98$ (solid line), $p=0.7$ (dashed line), $p=0.4$ (dashed-dotted line), $p=0$ (dotted line). All parameters are normalized to $\gamma_{1}$. }
\label{fig5}
\end{figure}

Figure~\ref{fig5} shows the results for the entanglement measure $\Upsilon$ as a function of $\Delta_{0}$ for various values of $p$. Since in the case of $p=0$, the creation of entanglement between the cavity modes was associated with a non-zero detuning, $\Delta_{L} \neq 0$, the role of SGC is illustrated most clearly if one assumes a resonant laser field. Consequently, we choose to limit our illustration of the creation of entanglement to a situation in which $\Delta_{L} =0$.

As before, for the case~\ref{sec3b}, the entanglement occurs for $p\neq 0$ and the optimal entanglement can be obtained at $\Delta_{0}=\Omega_{0}/2$. However, in contrast to the case A, the entanglement maximizes at $\Upsilon \approx -1$ for $\gamma_{2}=2\gamma_{1}$. It means that the entanglement maximizes when the transition rates of the dressed transition resonant with the undressed transition are equal.

In order to understand this behavior of entanglement, we consider the coefficients of the master equation in the limit of $\delta\gg\gamma_{i}$ and $\Omega_{0}\gg\gamma_{i}$ and find that in this limit, the master equation (\ref{e14}) reduces to the following form
\begin{align}
\frac{d}{dt}\rho_{c} =& -i\left(\delta_{12} +2\bar{A}\right)\left[ a_{1}^{\dagger }a_{1} +a_{2}^{\dagger }a_{2},\rho_{c}\right] \nonumber\\
&+i\left[\bar{D}a_{1}^{\dagger }a_{2}^{\dagger }+\bar{D}^{\ast}a_{1}a_{2},\rho_{c}\right] + L_{c}\rho_{c} ,\label{e32}
\end{align}
where
\begin{eqnarray}
\bar{A} &=& \frac{1}{4}g^{2}\left[\left(\frac{\sin^{4}\phi }{\Omega_{0}+\delta}+\frac{\cos^{4}\phi }{\Omega_{0}-\delta }\right)\left( \rho _{22}^{s}-\rho _{33}^{s}\right)\right. \nonumber\\
&+&\left. \frac{\sin^{2}\phi}{\Omega_{0}-\delta}\left( \rho _{11}^{s}-\rho_{33}^{s}\right) +
\frac{\cos^{2}\phi}{\delta}\left(\rho _{22}^{s}-\rho_{11}^{s}\right) \right] ,\nonumber\\
\bar{D} &=& \frac{\Omega_{0}{g^{2}}\sin \phi \cos^{2}\phi }{(\Omega_{0}-\delta)\delta }\rho _{12}^{s} ,\label{e33}
\end{eqnarray}
We may further simplify the master equation by choosing $\delta_{12} =-2\bar{A}$, which leaves only the non-linear term in its commutator part. Note that comparing to the case A, there is a qualitative difference in the dependence of the coefficient $\bar{D}$ on the density matrix elements. The magnitude of $\bar{D}$ depends now on the coherence between the states $\ket{\tilde 1}$ and $\ket{\tilde 2}$ but not on the population difference. The coherence is induced by spontaneous emission and can be different from zero only if $p\neq 0$. This means that the SGC is crucial for creation of entanglement between the cavity modes when the modes are coupled to different atomic transitions. As it is seen from Fig.~\ref{fig5}, the entanglement maximizes at $\Delta_{0}=\Omega_{0}/2$ and $p=1$. It is easy to show from Eqs.~(\ref{e24}) and (\ref{e26}) that for $\Delta_{0}=\Omega_{0}/2$ and $p=1$, in the steady state the population is trapped in the antisymmetric state $\ket a$. Thus, similar to the case A, the condition for the maximal entanglement coincides with the collapse of the atomic system into the pure trapping state. In this case, the coherence $\rho_{12}^{s} = -\alpha\beta$ and then the parameter $\bar{D}$ reduces to
\begin{eqnarray}
\bar{D} =  -\frac{\Omega_{0}g^{2}\sin^{2}2\phi}{4(\Omega_{0}-\delta)\delta}
\frac{\sqrt{\gamma_{1}\gamma_{2}}}{\gamma_{1} +\gamma_{2}\sin^{2}\phi} .\label{e34}
\end{eqnarray}
It is easily verified that the coefficient $\bar{D}$ attains its maximal value for $\phi=\pi/4$ and $\gamma_{2}=2\gamma_{1}$. Thus, the simple formula in Eq.~(\ref{e34}) predicts accurately the parameter values of the maximal entanglement in Fig.~\ref{fig5}.

\begin{figure}[hbt]
\includegraphics[width=\columnwidth,keepaspectratio,clip]{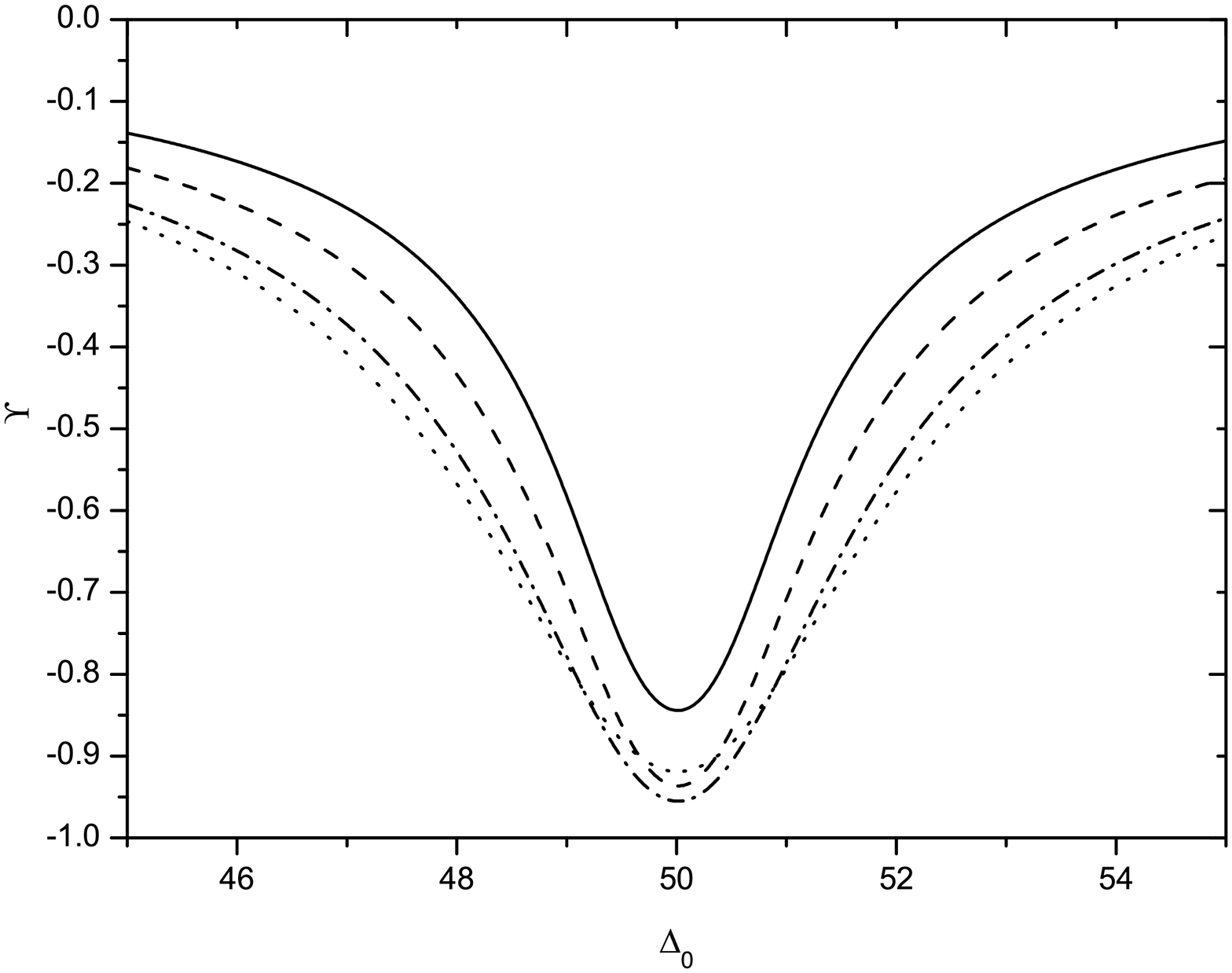}
\caption{The degree of entanglement $\Upsilon$ plotted as a function of $\Delta_{0}$ for the case of anti-parallel transition dipole moments, $p=-1$, with $\Delta_{L} = 0$, $\Omega = 50$, $\delta_{1} \approx\delta_{2} = 50$, $\kappa_{1} = \kappa_{2} = 0.72$, and different $\gamma_{2}/\gamma_{1}$:  $\gamma_{2}/\gamma_{1}=0.5$ (solid line), $\gamma_{2}/\gamma_{1}=1.0$ (dashed line), $\gamma_{2}/\gamma_{1}=2.0$ (dashed-dotted line), $\gamma_{2}/\gamma_{1}=3.0$ (dotted line). All parameters are normalized to~$\gamma_{1}$.}
\label{fig6}
\end{figure}

In concluding this section, we would like to point out that the qualitative features of entanglement between the cavity modes depend on whether the dipole moments of the atomic transitions are parallel $(p=1)$ or anti-parallel $(p=-1)$ to each other. We have already seen that in the case of parallel dipole moments and $\Delta_{0}=\Omega_{0}/2$, the population is trapped in the antisymmetric state irrespective of the laser detuning $\Delta_{L}$ and the ratio between the atomic spontaneous emission rates. However, for the anti-parallel dipole moments, the situation is different. It is not difficult to show from Eqs.~(\ref{e24}) and (\ref{e26}) that for $p=-1$ and $\Delta_{0}=\Omega_{0}/2$, the steady state populations of the states are
\begin{eqnarray}
\rho_{aa} =\left(\alpha^{2}-\beta^{2}\right)^{2}, \quad \rho_{ss} =4\alpha^{2}\beta^{2}, \quad \rho_{33}=0 ,\label{e29}
\end{eqnarray}
where $\alpha$ and $\beta$ are given in Eq.~(\ref{e27}). It is evident that in general the population is redistributed between the symmetric and antisymmetric states and only in the case of $\gamma_{1}=\gamma_{2}\sin^{2}\phi$ the population is trapped in one, the symmetric superposition state. A consequence of this population redistribution is the reduction of the entanglement between the cavity modes. This is shown in Fig.~\ref{fig6}, where we plot the entanglement measure for $p=-1$ and different ratios between the atomic spontaneous emission rates. For $\gamma_{2}\neq 2\gamma_{1}$, the magnitude of the entanglement is reduced and attains the maximal value of $\Upsilon =-1$ for $\gamma_{2}=2\gamma_{1}$. This is an another demonstration that the maximal entanglement between the modes is achieved only when two correlated atomic transitions decay rates obey $\gamma_{2}=2\gamma_{1}$.

\subsection{Other possible couplings of the modes to the atomic transitions}

Finally, we briefly comment on the other possible coupling configurations of the cavity modes to the atomic transitions. The two cases discussed above predict a large entanglement at practically the same conditions, with only different conditions imposed on the damping rates of the atomic transitions. An another possible configuration is to couple the cavity mode $\omega_{1}$ to the undriven transition $\ket 1 \leftrightarrow \ket 3$ and the mode $\omega_{2}$ to the driven transition $\ket 2 \leftrightarrow \ket 3$. One can see from Fig.~\ref{fig1b}, that this configuration is obtained from the case B simply by replacing $\delta$ by $-\delta$. Thus, a large entanglement could be generated in this configuration for the same condition as in the case B.

The most general configuration of the coupling constants is the case corresponding to all of the cavity modes simultaneously coupled to both atomic transitions. It is easily verified that this general case can be treated as a sum of two cases B with opposite detuning $\delta$. By combining the two cases together, we find that the magnitude of the effective coefficient $\bar{D}$ depends strongly on the sign of the parameter $p$. For $p=\pm 1$, the effective coefficient $\bar{D}$ takes the following form
\begin{eqnarray}
\bar{D}_{p=\pm 1} = \frac{g^{2}\Omega_{0}\sin^{2}2\phi}{2\left(\Omega _{0}^{2}- \delta^{2}\right)}
\frac{\sqrt{\gamma_{1}}\left(\sqrt{\gamma_{1}} \mp \sqrt{\gamma_{2}}\right)}{\gamma_{1}+\gamma_{2}\sin^{2}\phi} .\label{e35}
\end{eqnarray}
We see that depending on the sign of $p$ these two coupling configuration can interfere constructively or destructively resulting in an enhanced or reduced effective magnitude of the nonlinear process. For $p=1$ the configurations interfere destructively such that for $\gamma_{1}=\gamma_{2}$ the effective coefficient $\bar{D}$ vanish. On the other hand, for $p=-1$ the configurations interfere constructively which results in an enhanced amplitude of the nonlinear process. However, the resulting magnitude of the effective coefficient depends strongly on the ratio $\gamma_{2}/\gamma_{1}$ such that $\bar{D}$ is large for $\gamma_{2}/\gamma_{1}\ll 1$, but becomes very small, proportional to $\sqrt{\gamma_{1}/\gamma_{2}}$ in the opposite limit of $\gamma_{2}/\gamma_{1}\gg 1$. In other words, the three-level system can strongly entangle the cavity modes only if the spontaneous emission rate on the undriven transition is much larger than that of the driven transition.

We finish this section with a short discussion of a possibility to create entanglement between the cavity modes by the SGC in three-level atoms in the Lambda or cascade configurations. As we have shown, the crucial for the maximal entanglement is to trap the population in a pure superposition state of the atoms. However, it is well known that the SGC has a constructive effect on trapping of the population in a pure state only in the V-type atoms~\cite{fs04}. In the Lambda or cascade type atoms, the SGC has a destructive rather than constructive effect on the trapping phenomenon~\cite{ja92,ma98}.

The crucial for the entanglement is three-level atoms with parallel or nearly parallel dipole moments between the two atomic transitions. It is difficult in practice to find V-type systems with parallel or anti-parallel dipole moments. One of the possibility is to use sodium dimers, which can be modeled as a five-level molecule in which transitions with parallel and anti-parallel dipole moments can be selected~\cite{xy96,ww00}. An alternative solution is to engineer atomic systems with parallel dipole moments. For example, Zhou and Swain~\cite{zs00} showed that transitions with parallel dipole moments can be achieved in a three-level atom coupled to a cavity field with pre-selected polarization in the bad cavity limit. Agarwal~\cite{ag00} has demonstrated that an anisotropy in the vacuum can lead to quantum interference among the decay channels of close lying states. Another possibility is to align the dipole moments by a slow motion of the atoms through the medium~\cite{ne86}, or to apply a dc field to couple the upper levels of a three-level V-type atom with perpendicular dipole moments~\cite{zfs04}.

\section{Conclusions}\label{sec4}

We have proposed a scheme for generation on demand of a steady-state entanglement between two optical modes coupled to a V-type three-level atom. We have demonstrated that the condition for generation of the maximal entanglement between the modes is to create the complete population inversion between the dressed states of the coupling atomic system. In the case of a two-level atom composing the entangling atomic system, we have shown that the sufficient condition for entanglement between the modes is to create a population difference between dressed states of the driven atomic transition. However, we have found that the maximal entanglement cannot be created in this system since it is not possible to create the complete population inversion between the dressed states and at the same time maintain a strong coupling between the cavity modes mediated by the atom.
In the case of three-level atoms composing the entangling system, we have found that a stationary entanglement can be created on demand by tuning the Rabi frequency of the driving field to the difference between the atomic transition frequencies. The laser field mediates the spontaneously generated coherence between the atomic dipole transitions that allows to engineer the dissipation in such a way that the atoms evolve into a pure trapping state.

\section*{Acknowledgment}

This work is supported by the National Natural Science Foundation of China (Grant Nos. 60878004 and 11074087), the Ministry of Education under project SRFDP ( Grant No. 200805110002), and the Natural Science Foundation of Hubei Province.

\end{document}